\documentclass[%
reprint,
superscriptaddress,
amsmath,amssymb,
prb,
floatfix,
longbibliography
]{revtex4-2}

\usepackage{dcolumn}
\usepackage[pdftex]{color,graphicx}
\usepackage{bm}
\usepackage{floatrow}
\usepackage{hyperref}
\usepackage[draft]{todonotes} 
\usepackage{physics}
\usepackage{soul}
\usepackage{comment}
\usepackage{mathtools}
\usepackage{amsmath}

\newcommand{\ii}{\mathrm{i}}
\newcommand{\ee}{\mathrm{e}}
\newcommand{\dx}{\dd x}
\newcommand{\dy}{\dd y}

\begin{document}

\title{
Modal Purcell factor in $\mathcal{PT}$-symmetric waveguides
}

\author{Fyodor Morozko}
  \email{fyodormorozko95@gmail.com}
\author{Andrey Novitsky}
\affiliation{%
Department of Theoretical Physics and Astrophysics,
Belarusian State University, Nezavisimosti Avenue 4, 220030 Minsk, Belarus 
}
\author{Alina Karabchevsky}%
  \email{alinak@bgu.ac.il}
\affiliation{%
 School of Electrical and Computer Engineering, Ben-Gurion University, Beer-Sheva 8410501, Israel 
}

\date{\today}

\begin{abstract}
We study the spontaneous emission rate of a dipole emitter in $\mathcal{PT}$-symmetric environment of two coupled waveguides using the reciprocity approach generalized to non-orthogonal eigenmodes of non-Hermitian systems.
Considering emission to the guided modes, we define and calculate the modal Purcell factor composed of contributions of independent and interfering non-orthogonal modes leading to the emergence of cross-mode terms in the Purcell factor.
We reveal that the closed-form expression for the modal Purcell factor within the coupled mode theory slightly alters for the non-Hermitian coupled waveguide compared to the Hermitian case.
It is true even near the exceptional point, where the eigenmodes coalesce and the Petermann factor goes to infinity.
This result is fully confirmed by the numerical simulations of active and passive $\mathcal{PT}$-symmetric systems being the consequence of the mode non-orthogonality.
\end{abstract}

\maketitle

\section{\label{sec:intro}Introduction}
Quantum mechanics is based on the postulate that all physical observables must correspond to the real eigenvalues of quantum mechanical operators.
For a long time this assertion had been considered to be equivalent to the requirement of the Hermiticity of the operators.
The situation has changed after the seminal work \cite{ref:bender1998} of Bender and Boettcher, who discovered a wide class of non-Hermitian Hamiltonians exhibiting entirely real-valued spectra.
A number of intriguing properties are related to the non-Hermitian Hamiltonians possessing parity-time ($\mathcal{PT}$) symmetry that is the symmetry with respect to the simultaneous coordinate and time reversal. 
For instance, a system described by the Hamiltonian $\hat{H}=\frac{\hat{\vb{p}}^2}{2m} + V(\vb{r}) \neq \hat H^\dag$ is $\mathcal{PT}$-symmetric, if the complex potential $V(\vb{r})$ satisfies condition $V(\vb{r})=V^{*}(-\vb{r})$, where $^\dag$ and $^\ast$ stand for designation of the Hermitian and complex conjugations respectively.

A couple of important features of the $\mathcal{PT}$-symmetric Hamiltonians are worth mentioning~\cite{ref:elganainy2018,ref:zyablovsky2014,ref:wu2019}. First, their eigenfunctions corresponding to the real eigenvalues are not orthogonal.
Second, the systems are able to experience a phase transition from $\mathcal{PT}$-symmetric to $\mathcal{PT}$-symmetry-broken states, when system's parameters pass an exceptional point.
The transfer of the $\mathcal{PT}$ symmetry concept from quantum mechanics to optics is straightforward due to the similarity of the Schr\"odinger and diffraction equations \cite{ref:elganainy2007,ref:makris2008,ref:elganainy2018}.
Photonic $\mathcal{PT}$-symmetric structures are implemented by combining absorbing and amplifying spatial regions to ensure a complex refractive index $n(\vb{r})=n^*(-\vb{r})$ that substitutes the quantum-mechanical complex potential $V$.
A possibility of the experimental investigation of the $\mathcal{PT}$-symmetric structures certainly heats up the interest to this subject in optics \cite{ref:ruter2010a,ref:feng2013,Kremer2019} in order to apply these systems for sensing~\cite{Hodaei2017,Chen2017}, lasing, and coherent perfect absorption (anti-lasing) \cite{Sun2014,Wong2016}.

It was Purcell who revealed that a spontaneous emission rate is not an intrinsic property of the emitter, but is proportional to the local density of modes (density of photonic states) in the vicinity of the transition frequency~\cite{ref:purcell1946}.
In other words, the spontaneous emission rate is determined by an environment.
Phenomenon of the spontaneous emission enhancement owing to the influence of the environment is known now as the Purcell effect. The enhancement is defined as a ratio of the spontaneous emission rate in the system under consideration to that in the free space~\cite{gaponenko2010}.
With the development of nanotechnology, nanophotonics opens up new avenues for engineering spontaneous emission of quantum emitters in specific surrounding media~\cite{ref:klimov2001,ref:hughes2004,ref:anger2006,ref:kolchin2015,ref:karabchevsky2016,Su2019} including non-Hermitian media.
Investigation of the spontaneous emission of the dipole emitter inside a $\mathcal{PT}$-symmetric planar cavity has been recently performed by Akbarzadeh et~al in Ref.~\cite{ref:akbarzadeh2019}.
The authors have found suppression of the spontaneous relaxation rate of a two-level atom below the vacuum level.
A general theory of the spontaneous emission at the exceptional points of non-Hermitian systems was developed in Ref.~\cite{Pick2017} and revealed finite enhancement factors.

A number of methods including numerical techniques~\cite{ref:taflove2013} have been developed for calculation of the Purcell factor of dipole and quadrupole emitters in various environments.
The most general one is based on the calculation of Green's dyadics $\hat G({\bf r}, {\bf r}_0)$. Since the photonic local density of states is proportional to the imaginary part of the dyadic ${\rm Im}\hat G({\bf r}_0, {\bf r}_0)$~\cite{ref:novotny2012}, the purely quantum phenomenon of spontaneous emission can be reduced to the problem of classical electrodynamics. The Purcell factor $F_p = P/P_0$ can be written in terms of the powers $P$ and $P_0$ emitted by a source in an environment and in the free space, respectively.
This approach is widely adopted and can be exploited, e.g., for description of the spontaneous relaxation of molecules in absorbing planar cavities~\cite{ref:tomas1997}, explanation of the surface-enhanced Raman scattering \cite{Maslovski2019}, finding anomalous Purcell factor scaling in hyperbolic metamaterials~\cite{Wang2019}, and many others.

The Purcell factor can be calculated separately for each of the discrete scattering channels.
Due to the highly demanding field of photonic integrated circuitry (PIC) offering chip-scale miniaturization of actual devices and transformation of academy governed knowledge to the industry, recently the research has been accelerated towards utilization of important optical phenomena in integrated photonic devices as summarised in the recent Review on on-chip nanophotonics and future challenges~\cite{ref:karabchevsky2020}.
For instance, just a couple of years ago, the modal Purcell factor for the basic element of PIC planar waveguide was introduced within the scattering matrix formalism~\cite{ref:ivanov2017}.
A year after, another approach based on application of the reciprocity theorem was developed and successfully exploited in a ring resonator configuration~\cite{ref:schulz2018}.

Here, we generalize the reciprocity-theorem formalism to the case of non-Hermitian systems with non-orthogonal modes and
define the modal Purcell factor for a point-source emitter placed in the vicinity of such systems.
We examine the developed theory by studying the influence of the non-Hermiticity and the non-orthogonality on the spontaneous
emission rate of the point-source emitter placed near the coupled-$\mathcal{PT}$-symmetric waveguide systems. We show
analytically, utilizing the coupled mode approach and verify numerically using Finite-difference frequency-domain (FDFD based mode solver,
that although $\mathcal{PT}$-symmetric systems
are known to exhibit Purcell factor enhancement near exceptional point as reported in~\cite{Pick2017},
in principle no change in modal Purcell factor occurs for $\mathcal{PT}$-symmetric coupled-waveguides system
even near exceptional point where the supermodes coalesce leading to infinite values of the Petermann factor.

\begin{figure}[ht]
  \centering
  \includegraphics[width=\linewidth]{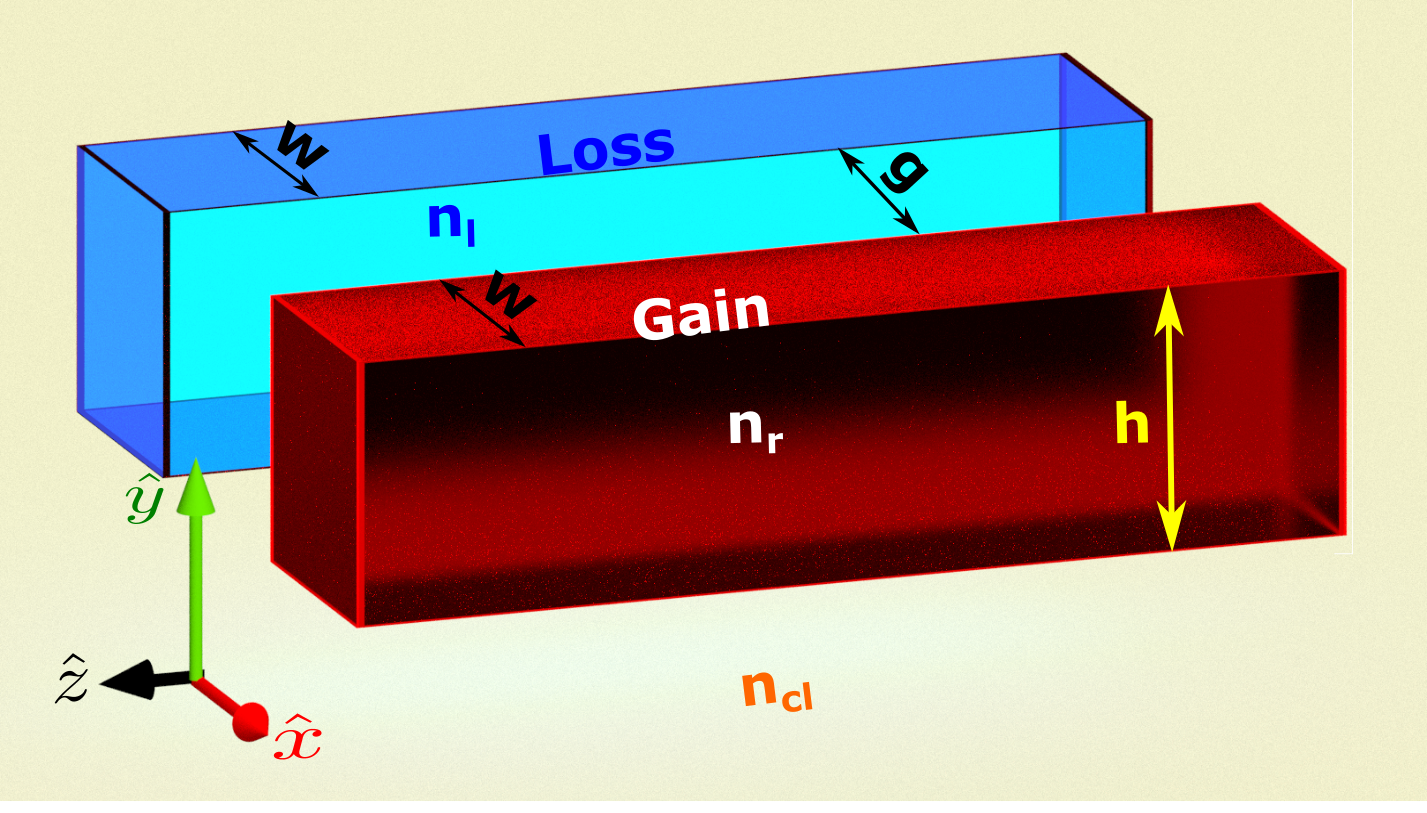}
  \caption{Schematics of the $\mathcal{PT}$-symmetric system: Gain (refractive index $n_r=n_{\rm{co}} + \ii\gamma$)
    and Loss ($n_1=n_{\rm{co}} - \ii\gamma$) waveguides of the width $w$ and height $h$ are embedded in the dielectric medium
    with index of $n_{cl}$.
    Waveguides are separated by the distance $g$.
    Modes propagate in the $\hat{z}$ direction.}
  \label{fig:wg}
\end{figure}

The rest of the paper is organized in the following way.
In Section~\ref{sec:method}, we formulate a method for the Purcell factor calculation based on the reciprocity approach that accounts for the modes non-orthogonality.
In Section~\ref{sec:cmt}, we probe the developed formalism by considering a $\mathcal{PT}$-symmetric coupled waveguides
system in terms of coupled mode approach and reveal no dependence of the modal Purcell factor on the non-Hermiticity.
In Section~\ref{sec:results}, we show the proof-of-concept calculations of the Purcell factor for the system demonstrated in Fig.~\ref{fig:wg}
and reveal an agreement with the results obtained using the coupled-mode approach.
Eventually, the Section V concludes the paper.

\section{\label{sec:method} Modal Purcell factor for non-Hermitian waveguides}

\subsection{Reciprocity approach}
Utilizing the reciprocity approach (a method for calculating the power $P$ emitted by a current source into a particular propagating mode leaving an open optical system), we normalize this power by the power of radiation into the free space $P_0$ to find the so called \textit{modal Purcell factor}~$F_p$.

We consider an emitting current source (current density distribution $\vb{J}_1$) situated inside a coupled waveguide system with two exit ports at $z_1$ and $z_n$ ~\cite{ref:schulz2018}.
For brevity, we introduce a 4-component vector joining transverse electric and magnetic fields as
\begin{equation}
  \ket{\psi(z)}=\mqty(\vb{E}_{t}(x,y,z) \\ \vb{H}_{t}(x,y,z)).
  \label{eq:dirac_notation}
\end{equation}
In this way we can describe the fields of guiding (and leaking) modes.
For the $i$th mode we write
\begin{equation}
  \ket{M_i(z)}=\mqty(\vb{E}_{t,i}(x, y, z) \\ \vb{H}_{t,i}(x, y, z))=\ket{i}\ee^{-\ii\beta_iz},
  \label{eq:Mode_i}
\end{equation}
where
\begin{equation}
  \ket{i}=\mqty(\vb{e}_{t,i}(x, y) \\ \vb{h}_{t,i}(x, y))
  \label{eq:mode_i}
\end{equation}
and
\begin{equation}
  \mqty(\vb{E}_{t,i}(x, y, z) \\ \vb{H}_{t,i}(x, y, z)) =
  \mqty(\vb{e}_{t,i}(x, y) \\ \vb{h}_{t,i}(x, y))\ee^{-\ii\beta_iz}.
\end{equation}

Here we define the inner product as a cross product of the bra-electric and ket-magnetic fields integrated over the cross-section $z={\rm const}$:
\begin{equation}
  \braket{\phi_1}{\phi_2} \equiv \intop_{}
  \left( \vb{E}_{1}\times\vb{H}_{2} \right) \cdot \vu{z} \dx\dy
  \label{eq:inner_product}
\end{equation}
Such a definition is justified by the non-Hermitian system we explore.
In the above and following relations we can drop $t$ subscripts because $z$
component of the vector products depends only on transverse components.
It is well known that the modes of Hermitian systems are orthogonal in the sense
\begin{equation}
  \intop \left( \vb{e}_{i}\times\vb{h}_{j}^{*} \right) \cdot \vu{z}\dx\dy \sim \delta_{ij},
  \label{eq:orth_conventional}
\end{equation}
where $\delta_{ij}$ is the Kronecker delta.
However, the loss and gain channels of the non-Hermitian waveguide break the orthogonality of the modes.
In this case, one should use a non-conjugate inner product~\cite{ref:snyder1984,ref:svendsen2013,ref:wu2019} bringing us to the orthogonality relationship
\begin{equation}
  \braket{i}{j} = \intop \left( \vb{e}_{i}\times\vb{h}_{j} \right) \cdot \vu{z} \dd x \dd y = 2 N_i \delta_{ij},
  \label{eq:orth}
\end{equation}
where $N_i$ is a normalization parameter.
It worth noting, that redefinition of the inner product is required in non-Hermitian quantum mechanics.
It appears that left and right eigenvectors of non-Hermitian operators obey the so-called biorthogonality relations.
Discussion of quantum mechanics based on biorthogonal states is given in
\cite{ref:weigert2003,ref:mostafazadeh2010,ref:moiseyev2011a,ref:brody2016}.

The fields excited by the current source $\vb{J}_1$ at the cross-section of exit ports can be expanded into a set of modes as follows
\begin{align}
  \ket{\psi_1(z_1)}&=\sum_{i} A_{i,z_1} \ket{i,z_1}, \nonumber \\
  \ket{\psi_1(z_n)}&=\sum_{i} A_{-i,z_n} \ket{-i,z_n}.
\label{eq:expansion}
\end{align}
Here $A_{i,z_1}$ and $A_{-i,z_n}$ are the amplitudes of the modes propagating forward to port $z_1$ and backward to port $z_n$, respectively,
$\ket{i,z_1}$, $\ket{-i,z_n}$ are respectively eigenmodes of ports $z_1$ and $z_n$ propagating
from the cavity.

In our notations the Lorentz reciprocity theorem
\begin{equation}
  \intop_{\delta V} \left( \vb{E}_{1}\times\vb{H}_{2} - \vb{E}_{2}\times\vb{H}_{1} \right)
  \cdot \vu{z} \dd x \dd y
  = \intop_{V} \left(\vb{E}_{2} \cdot \vb{J}_{1} - \vb{E}_{1} \cdot \vb{J}_{2} \right) \dd V.
  \label{eq:reciprocity}
\end{equation}
should be rewritten as
\begin{multline}
\braket{\psi_1(z_1)}{\psi_2(z_1)}-\braket{\psi_2(z_1)}{\psi_1(z_1)}\\
-\braket{\psi_1(z_n)}{\psi_2(z_n)}+\braket{\psi_2(z_n)}{\psi_1(z_n)}\\
  = \intop_{V}\left( \vb{E}_{2}\cdot\vb{J}_{1}-\vb{E}_{1}\cdot\vb{J}_{2} \right)\dd V,
  \label{eq:reciprocity1}
\end{multline}
where $\delta V$ is the surface enclosing the cavity volume $V$ between two planes $z=z_1$ and $z=z_n$.
In Eq.~\eqref{eq:reciprocity1}, ${\bf J}_1$ and $\ket{\psi_1}$ are defined above, while the source ${\bf J}_2$ and the fields $\ket{\psi_2}$ produced by it can be chosen as we need.
Let the source current ${\bf J}_2$, being outside the volume $V$ (${\bf J}_2 = 0$), excites a single mode $\ket{-k,z_1}$.
In general, this mode is scattered by the cavity $V$ and creates the set of transmitted and reflected modes as discussed in \cite{ref:schulz2018}.
In our case the cavity is a tiny volume ($z_1\approx z_n$) of the waveguide embracing the source ${\bf J}_1$.
Therefore, the field just passes the waveguide without reflection and we get
\begin{align}
  \ket{\psi_{2}(z_1)} &= B_{-k, z_1} \ket{-k,z_1},
  \label{eq:psi2z1}\\
  \ket{\psi_{2}(z_n)} &= B_{-k, z_n} \ket{-k,z_n}.
  \label{eq:psi2zn}
\end{align}

Forward and backward transverse modal fields $\vb{e}_{t,i}$ and $\vb{h}_{t,i}$
($\vb{e}_{t,i}\vu{z} = 0$) used in Eq.~(\ref{eq:reciprocity1}) satisfy the symmetry relations
\begin{equation}
  \vb{e}_{t,-i}=\vb{e}_{t,i}, \qquad\vb{h}_{t,-i}=-\vb{h}_{t,i}
  \label{eq:forward_backward}
\end{equation}
both in the case of Hermitian and non-Hermitian ports.

This means that the inner product of modes also meets the symmetry relations for its bra- and ket-parts: $\braket{i}{j}=\braket{-i}{j}$ and $\braket{i}{j}=-\braket{i}{-j}$.
Adding the orthogonality conditions~(\ref{eq:orth}), one straightforwardly derives
\begin{align}
\braket{\psi_1(z_1)}{\psi_2(z_1)} &= - \braket{\psi_2(z_1)}{\psi_1(z_1)} = - 2 A_{k,z_1} B_{-k,z_1} N_{k,z_1},  \nonumber \\
\braket{\psi_1(z_n)}{\psi_2(z_n)} &= \braket{\psi_2(z_n)}{\psi_1(z_n)} = 0,
\end{align}
where $N_{k,z_1}$ the norm of the mode $\ket{k,z_1}$ as defined in \eqref{eq:orth}.
These inner products in the general case of reflection and transmission of the reciprocal mode by a cavity are given in Appendix.

By substituting these equations into Eq.~\eqref{eq:reciprocity1}, we arrive at the amplitude $A_{k,z_1}$ of the mode excited by the source current $\vb{J}_1$ 
\begin{equation}
  A_{k, z_1} = -\frac{1}{4 B_{-k, z_1} N_{k, z_1}} 
  \intop_{V} \vb{E}_{2,-k} \cdot \vb{J}_{1} \dd V,
  \label{eq:minus4ab}
\end{equation}
where $\vb{E}_{2,-k} = B_{-k, z_1}\vb{e}_{-k}(x,y)\ee^{\ii \beta_k (z-z_1)}$ is the electric field created by the excitation of the system with reciprocal mode $\ket{-k,z_1}$ at the port $z_1$.

\subsection{Purcell factor}
As an emitter we consider a point dipole oscillating at the circular frequency $\omega$ and having the current density distribution
\begin{equation}
  \vb{J}_{1}\left( \vb{r} \right) =
  \ii\omega\vb{p}\delta\left( \vb{r}-\vb{r}_0 \right),
  \label{eq:dipole_current}
\end{equation}
where $\vb{p}$ is the dipole moment of the emitter and $\vb{r}_0$ is its position.
Then we are able to carry out the integration in Eq.~\eqref{eq:minus4ab} and obtain
\begin{equation}
  A_{k,z_1}=-\frac{\ii\omega}{4 B_{-k, z_1}N_{k,z_1}} \vb{E}_{2,-k}
  \left( \vb{r}_0 \right) \cdot \vb{p}.
  \label{eq:Akz1}
\end{equation}
Here we observe a dramatic difference compared to the Hermitian case considered in Ref.~\cite{ref:schulz2018}. This difference appears due to the fact that now the expansion coefficients $A_{k, z_1}$ are not directly related to the powers carried by the modes. Finding a power carried by a specific mode is a challenge. To circumvent this challenge, we propose a calculation of the total power carried by the set of modes as we describe below.

The power emitted by the current source $\vb{J}_1$ into the port $z_1$ can be written as
\begin{equation}
  P = \frac{1}{2} \mathrm{Re}\intop_{z=z_1} \left( \vb{E}_{1} \times \vb{H}^{*}_{1} \right) \cdot
  \vu{z} \dd x \dd y = \frac{1}{2} \braket{\psi_1(z_1)}{\psi_1^\ast(z_1)},
  \label{eq:power1}
\end{equation}
where $\ket{\psi^\ast}=(\vb{E}_t^*,\vb{H}_t^*)^T$.
Expanding the electromagnetic fields $\ket{\psi_1(z_1)}$ according to Eq.~(\ref{eq:expansion})
we represent the power transmitted through the port Eq. (\ref{eq:power1}) as follows
\begin{equation}
  P = \mathrm{Re}\sum_{k,l}A_{k, z_1} A^{*}_{l, z_1} P_{kl},
  \label{eq:power1_expand}
\end{equation}
where $P_{kl}$ is the so called cross-power equal to the Hermitian inner product of the modal fields
\begin{equation}
  P_{kl,z_1} = \frac{1}{2} \braket{k,z_1}{l,z_1^*} =
  \frac{1}{2}\intop_{z=z1}\left( \vb{e}_{k,z_1} \times \vb{h}^{*}_{l,z_1} \right)\vdot\vu{z}\dx\dy.
  \label{eq:cross_power}
\end{equation}
For $k = l$ the cross-power reduces to the mode power $P_k = P_{kk}$. By considering the expansion coefficients \eqref{eq:Akz1} we rewrite the power (\ref{eq:power1_expand}) in terms of the reciprocal fields $\vb{E}_{2,-k}$ as
\begin{multline}
  P = \frac{\omega^{2}}{16}
    \Re
    \sum_{k,l}
    \frac{
  ( \vb{E}_{2,-k}\left( \vb{r}_0 \right)\cdot \vb{p} )
  ( \vb{E}_{2,-l}^{*}\left( \vb{r}_0 \right)\cdot \vb{p}^\ast )
    }
    {B_{-k}B_{-l}^{*} N_{k} N_{l}^*}
    P_{kl}
    \\
  =\frac{\omega^{2}}{16}
    \Re
    \sum_{k,l}
    \frac{
  (\vb{e}_{-k}\left( x_0, y_0 \right)\cdot \vb{p} )
  ( \vb{e}_{-l}^{*}\left( x_0, y_0 \right)\cdot \vb{p}^\ast )
    }
    {N_{k} N_{l}^*}
    P_{kl}.
    \label{eq:power_final}
\end{multline}
The last equality is the consequence of the substitution of $\vb{E}_{2,-k}$ at the emitter position ${\bf r}_0 = (x_0, y_0, z_0)$ and taking into account negligible dimensions of the cavity $z_1\approx z_n\approx z_0$.
Note that here we dropped $z_1$ subscripts.

In order to find the Purcell factor we divide Eq. (\ref{eq:power_final}) by the power emitted by the same dipole into the free space
\begin{equation}
  P_{0} = \frac{\mu_{0}}{12\pi c} \omega^{4} |p|^{2},
  \label{eq:P0}
\end{equation}
where $\mu_0$ is the vacuum permeability and $c$ is the speed of light in vacuum.
The dipole moment, located in the $xy$ plane, can be presented using the unit vector $\hat {\bf p}$ as follows
\begin{equation}
 \vb{p}=p\vu{p},
\end{equation}
therefore,
\begin{equation}
  \vb{E}_{2,-k}(\vb{r}_0)\cdot\vb{p}=\vb{E}_{2,-k}(\vb{r}_0)\cdot\vu{p}p=E_{p,k}(\vb{r}_0)p.
  \label{eq:Edotp}
\end{equation}
Here $E_{p,k}$ denotes projection of the vector $\vb{E}_{2,-k}$ onto the dipole orientation vector $\vu{p}$
\begin{equation}
  E_{p,k}=\vb{E}_{2,-k}\cdot\vu{p}.
  \label{eq:cos_alpha}
\end{equation}
Then the Purcell factor reads
\begin{equation}
  F_{p} = \frac{P}{P_0}
  =\frac{3\pi c}{4\omega^{2} \mu_{0}} \mathrm{Re}
  \sum_{k,l}
    \frac{
      e_{p,k}\left( x_0, y_0 \right)
      e_{p,l}^{*}\left( x_0, y_0 \right)
    }{N_{k} N_{l}^*}P_{kl}.
  \label{eq:Fpurcell}
\end{equation}

It is convenient to rewrite Eq. \eqref{eq:Fpurcell} through the normalized fields as
\begin{equation}
  F_{p}=
  \frac{3\pi c}{4\omega^{2} \mu_{0}}\sum_{kl}\hat{e}_{p,k}\hat{e}^*_{p,l}K_{kl}\hat{P}_{kl},
  \label{eq:Fp_normed}
\end{equation}
where we have introduced power-normalized modal electric fields
\begin{equation}
  \hat{\vb{e}}_{2, i} = \frac{\vb{e}_{2,i}}{\sqrt{P_{i}}}
  \label{eq:e_power_normalized}
\end{equation}
and normalized cross-power coefficients
\begin{equation}
  \hat{P}_{kl} = \frac{1}{\sqrt{P_kP_l}}P_{kl}.
\end{equation}
Here we generalize the well-known Petermann factor~\cite{ref:petermann1979}
\begin{equation}
  K_i = K_{ii}
  \label{eq:Petermann_factor}
\end{equation}
defining cross-mode Petermann factor
\begin{equation}
  K_{kl} = \frac{P_k}{N_k}\frac{P_l}{N_l^*}=
  \frac{\Re\braket{k}{k^*}}{\braket{k}}\frac{\Re\braket{l}{l^*}}{\braket{l}^*}.
  \label{eq:cross_Petermann_factor}
\end{equation}
It should be noticed that the Petermann factor is often related to the mode non-orthogonality~\cite{Pick2017,ref:siegman1989,ref:berry2003,ref:yoo2011} being obviously equal to the unity for Hermitian systems owing to the coincidence of the norm $N_i$ and power $P_i$ in this case.
The modal Purcell factor can be naturally divided into two parts, the first of which is the sum of all diagonal $(k=l)$ terms, while the second part is the sum of non-diagonal $(k\neq l)$ terms:
\begin{equation}
  F_p = F_{p, \mathrm{diag}} + F_{p,\mathrm{non-diag}} =
  \sum_kF_{p, k}+\sum_{k\neq l}F_{p, kl},
  \label{eq:Fp_two_terms}
\end{equation}
where
\begin{align}
  F_{p, i}&=\frac{3\pi c}{4\omega^{2} \mu_{0}}
    \abs{\hat{e}_{p,k}}^2 K_{i},
  \label{eq:Fp_diag}\\
  F_{p, kl}&=\frac{3\pi c}{4\omega^{2} \mu_{0}}
  \hat{e}_{p,k}\hat{e}^*_{p,l} K_{kl}\hat{P}_{kl}.
  \label{eq:Fp_non-diag}
\end{align}
In the Hermitian case, the non-diagonal terms \eqref{eq:Fp_non-diag} reduce to zero due to the regular orthogonality of the modes expressed by $\hat{P}_{kl}=\delta_{kl}$.
That is why the Purcell factor \eqref{eq:Fp_normed} applied to Hermitian systems coincides with the expression in Ref.~\cite{ref:schulz2018}.

\section{\label{sec:cmt} Modal Purcell factor within the Coupled Mode Theory}

To get some insight on the behavior of the modal Purcell factor, let us analyze the system of two coupled waveguides using the coupled mode theory as adopted in $\mathcal{PT}$-symmetry related literature.
We express the total field at the port $z_1$ in the coupled system in terms of the modes
$\ket{g}$ and $\ket{l}$ of isolated gain and loss waveguides
with corresponding $z$-dependent amplitudes $g$ and $l$ as
\begin{equation}
  \ket{\psi_1}=g(z)\ket{g}+l(z)\ket{l}.
  \label{eq:gl_ansatz}
\end{equation}
We assume the overlap between the modes of isolated waveguides is negligible
(weak coupling condition), therefore, the modes are orthogonal and normalized as follows
\begin{align}
  \braket{g}{l}&=\braket{g}{l^*}=0,\\
  \braket{g}&=\braket{l}=1.
  \label{eq:gl_norm}
\end{align}
One more assumption is introduced for the sake of simplicity:
\begin{equation}
  \braket{g}{g^*}=\braket{l}{l^*}=1.
\end{equation}
It implies that the Hermitian norms of the isolated modes
are equal to the non-Hermitian norms or, in other words,
the Petermann factors for the modes equal unity.

$\mathcal{PT}$ operator converts the mode of isolated lossy waveguide to the mode of the isolated gain waveguide and vice versa that is
\begin{subequations}
\begin{align}
  \mathcal{PT}\ket{g} &= \ket{l},\\
  \mathcal{PT}\ket{l} &= \ket{g}.
\end{align}
\label{eq:PTgl}
\end{subequations}
Spatial evolution of amplitudes is governed by the system of coupled equations
\begin{equation}
  \ii\dv{z}\mqty[g\\l] =
  \mqty[\Re(\beta+\delta)-\ii\alpha/2&&\kappa\\
    \kappa&&\Re(\beta+\delta)+\ii\alpha/2]\mqty[g\\l]
\end{equation}
where $\beta$ is a propagation constant, $\kappa$ is a coupling coefficient,
$\delta$ is a correction to the propagation constant, $\alpha$ is an effective gain (or loss).
It can be shown that due to the weak coupling and relations \eqref{eq:PTgl}
the coupling constant $\kappa$ is real \cite{ref:chuang1987,ref:elganainy2007}.

\subsection{$\mathcal{PT}$-symmetric regime}
In $\mathcal{PT}$-symmetric regime, the system has the supermodes of the form
\begin{equation}
  \ket{1,2}=\ket{g}\pm\ee^{\pm\ii\theta}\ket{l}
  \label{eq:eigmodes_sym}
\end{equation}
with corresponding eigenvalues
\begin{equation}
  \beta_{1,2}=\Re(\beta+\delta)\pm\kappa\cos\theta,
  \label{eq:eigvals_sym}
\end{equation}
where $\sin\theta=\alpha/2\kappa$.

To find the modal Purcell factor in terms of coupled modes we substitute
the modes in the form \eqref{eq:eigmodes_sym}
into expression \eqref{eq:Fp_normed}.

Then the quantities $K_{kl}$ and $\hat{P}_{kl}$ can be written in the closed form as
\begin{subequations}
\begin{align}
  K_{1}=\frac{\Re\braket{1}{1^*}^2}{\abs{\braket{1}}^2}
  &=\frac{2}{1+\cos2\theta},\\
  K_{2}=\frac{\Re\braket{2}{2^*}^2}{\abs{\braket{2}}^2}
  &=\frac{2}{1+\cos2\theta},\\
  K_{12}=\frac{\Re\braket{1}{1^*}}{\braket{1}}
  \frac{\Re\braket{2}{2^*}}{\braket{2}^*}
  &=\frac{2(1+\ee^{-\ii2\theta})^2}{(1+\cos2\theta)^2},\\
  K_{21}=\frac{\Re\braket{2}{2^*}}{\braket{2}}
  \frac{\Re\braket{1}{1^*}}{\braket{2}^*}
  &=\frac{2(1+\ee^{+\ii2\theta})^2}{(1+\cos2\theta)^2},
\end{align}
\label{eq:K_kl_sym}
\end{subequations}
\begin{subequations}
\begin{align}
  \hat{P}_{12}
  =\frac{\braket{1}{2^*}}{\sqrt{\braket{1}{1^*}\braket{2}{2^*}}}
  &=\frac{1}{2}(1-\ee^{\ii2\theta}),\\
  \hat{P}_{21}
  =\frac{\braket{2}{1^*}}{\sqrt{\braket{1}{1^*}\braket{2}{2^*}}}
  &=\frac{1}{2}(1-\ee^{-\ii2\theta}).
\end{align}
\label{eq:P_kl_sym}
\end{subequations}
Normalized field projections $\hat{e}_{p,k}$ in
the basis of isolated modes read
\begin{subequations}
\begin{align}
  \hat{e}_{p,1}&=\frac{1}{\sqrt{\frac12\bra{1}\ket{1^*}}}(\hat{e}_{p,g}+\ee^{\ii\theta}\hat{e}_{p,l})
  =\hat{e}_{p,g}+\ee^{\ii\theta}\hat{e}_{p,l},\\
  \hat{e}_{p,2}&=\frac{1}{\sqrt{\frac12\bra{2}\ket{2^*}}}(\hat{e}_{p,g}-\ee^{-\ii\theta}\hat{e}_{p,l})
  =\hat{e}_{p,g}-\ee^{-\ii\theta}\hat{e}_{p,l}.
  \label{eq:ep12_sym}
\end{align}
\end{subequations}
In above expressions $\hat{e}_{p,g}$ and $\hat{e}_{p,l}$ denote projections of the fields of
backward-propagating isolated modes onto dipole orientation.
If the emitter dipole moment is perpendicular to $\hat{z}$,
projections of backward-propagating modal fields are equal to
projections of forward-propagating ones.

Performing calculation of the modal Purcell factor \eqref{eq:Fp_normed} using
relations (\ref{eq:K_kl_sym}-\ref{eq:P_kl_sym}) we obtain
\begin{equation}
  F_{p}=F_{p,\mathrm{diag}}+F_{p,\mathrm{non-diag}}=
  \frac{6\pi c}{\omega^{2} \mu_{0}}
  (\abs{\hat{e}_{p,g}}^2+\abs{\hat{e}_{p,l}}^2).
  \label{eq:Fp_sym}
\end{equation}
Diagonal and non-diagonal terms separately take the form
\begin{subequations}
\begin{align}
  F_{p,\mathrm{diag}}&=\frac{3\pi c}{4\omega^{2} \mu_{0}}
    \frac{4}{1+\cos2\theta}(\abs{\hat{e}_{p,g}}^2+\abs{\hat{e}_{p,l}}^2),\\
  F_{p,\mathrm{non-diag}}&=-\frac{3\pi c}{4\omega^{2} \mu_{0}}
  \frac{2(1-\cos2\theta)}{1+\cos2\theta}(\abs{\hat{e}_{p,g}}^2+\abs{\hat{e}_{p,l}}^2).
\end{align}
\label{eq:Fp_terms_sym}
\end{subequations}

It is curious that although both diagonal and non-diagonal terms \eqref{eq:Fp_terms_sym}
are singular at the EP corresponding to $\theta_{EP}=\pi/2$ and $\cos2\theta_{EP}=-1$,
the singularities cancel each other making the modal Purcell factor finite and independent of $\theta$.
The modal Purcell factor (\ref{eq:Fp_sym}) depends solely on the mode profiles of the isolated modes in $\mathcal{PT}$-symmetric regime.

Further we will show
that the similar conclusion holds, when the $\mathcal{PT}$ symmetry is violated.

\subsection{Broken $\mathcal{PT}$ symmetry regime}

In the $\mathcal{PT}$-broken regime, supermodes of the system of coupled waveguides take the form
\begin{equation}
  \ket{1,2}=\ket{g}+\ii\ee^{\mp\theta}\ket{l},
  \label{eq:eigmodes_broken}
\end{equation}
while eigenvalues read
\begin{equation}
  \beta_{1,2}=\Re(\beta+\delta)\pm\ii\kappa\sinh\theta,
  \label{eq:eigvals_broken}
\end{equation}
where $\cosh\theta=\alpha/2\kappa$.

Calculating
\begin{subequations}
\begin{align}
  K_{1}&=\coth^2{\theta},\\
  K_{2}&=\coth^2{\theta},\\
  K_{12}&=-\coth^2{\theta},\\
  K_{21}&=-\coth^2{\theta},
\end{align}
\end{subequations}
\begin{subequations}
\begin{align}
  \hat{P}_{12}&=\frac{1}{\cosh\theta},\\
  \hat{P}_{21}&=\frac{1}{\cosh\theta},
\end{align}
\end{subequations}

\begin{subequations}
\begin{align}
  \hat{e}_{p,1}&=
  \frac{1}{\sqrt{\frac12(1+\ee^{-2\theta})}}(\hat{e}_{p,g}+\ii\ee^{-\theta}\hat{e}_{p,l}),\\
  \hat{e}_{p,2}&=
  \frac{1}{\sqrt{\frac12(1+\ee^{2\theta})}}(\hat{e}_{p,g}+\ii\ee^{\theta}\hat{e}_{p,l})
\end{align}
\label{eq:ep12_broken}
\end{subequations}
we straightforwardly derive the diagonal and non-diagonal terms
\begin{subequations}
\begin{multline}
  F_{p,\mathrm{diag}}=\\
  \frac{3\pi c}{4\omega^{2} \mu_{0}}
  \frac{2\cosh\theta}{\sinh^2\theta}
  \left((\abs{\hat{e}_{p,g}}^2+\abs{\hat{e}_{p,l}}^2)\cosh\theta-
  2\Im(\hat{e}_{p,g}^*\hat{e}_{p,l})\right),
\end{multline}
\begin{multline}
  F_{p,\mathrm{non-diag}}=\\
  -\frac{3\pi c}{4\omega^{2} \mu_{0}}
  \frac{2}{\sinh^2\theta}
  \left(\abs{\hat{e}_{p,g}}^2+\abs{\hat{e}_{p,l}}^2-
  2\cosh\theta\Im(\hat{e}_{p,g}^*\hat{e}_{p,l})\right)
\end{multline}
\end{subequations}
as well as the modal Purcell factor
\begin{equation}
  F_{p}=F_{p,\mathrm{diag}}+F_{p,\mathrm{non-diag}}=
  \frac{6\pi c}{\omega^{2} \mu_{0}}
  (\abs{\hat{e}_{p,g}}^2+\abs{\hat{e}_{p,l}}^2).
  \label{eq:Fp_broken}
\end{equation}

The main result of this section is that although diagonal and non-diagonal terms
of the modal Purcell factor diverge at the EP,
the modal Purcell factor itself does not exhibit a singular behavior
when approaching to the EP either from the left or right side.

Though we do not carry out a rigorous analysis of the behavior at the EP
accounting for the degeneracy of the modes as it was done in Ref.~\cite{Pick2017}, the developed approach leads to the well-defined expressions \eqref{eq:Fp_sym}~and~\eqref{eq:Fp_broken} for $F_p$ at the exceptional point.

\section{\label{sec:results}Numerical Results}
In this section we probe the theory developed in the previous section
by analyzing numerically an optical system consisting of two coupled rectangular waveguides separated by the distance
$g$ as schematically shown in Fig.~\ref{fig:wg}.
Complex refractive indices of the left (Loss) and right (Gain) waveguides are equal to $n_1=n_{\rm{co}} - \ii\gamma$ and $n_r=n_{\rm{co}} + \ii\gamma$ respectively to satisfy $\mathcal{PT}$-symmetry condition $n(x) = n^\ast(-x)$, where $n_{\rm{co}}$ is the refractive index and $\gamma>0$ is the gain/loss (non-Hermiticity) parameter.
The waveguides are embedded in the transparent ambient medium with refractive index $n_{\rm{cl}}$. Light propagates in the $z$-direction.

\begin{figure}[t!b!]
  \includegraphics[width=\linewidth]{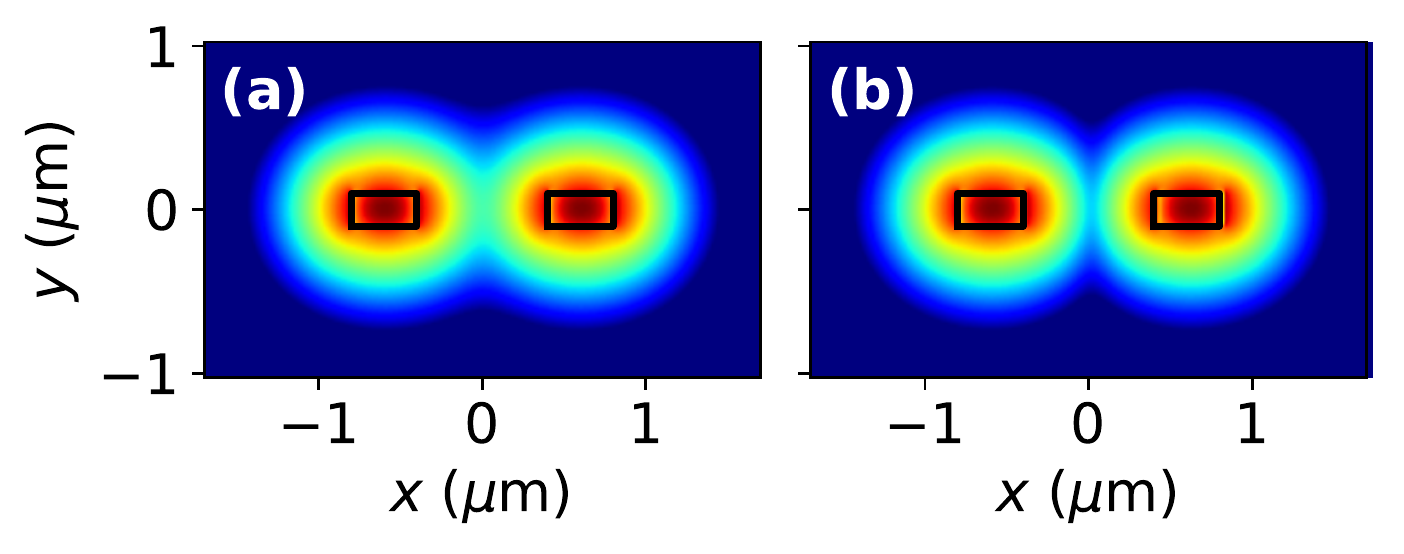}
  \caption{
    Distribution of the electric field component $E_x$ in the
    $\mathcal{PT}$-symmetric regime ($\gamma=3.5\times10^{-4}$).
    Distributions of $\abs{E_x}$ for (a) ``even'' and (b) ``odd'' supermodes.
  }
  \label{fig:modes_below_ep}
\end{figure}

\begin{figure}[t!b!]
  \includegraphics[width=\linewidth]{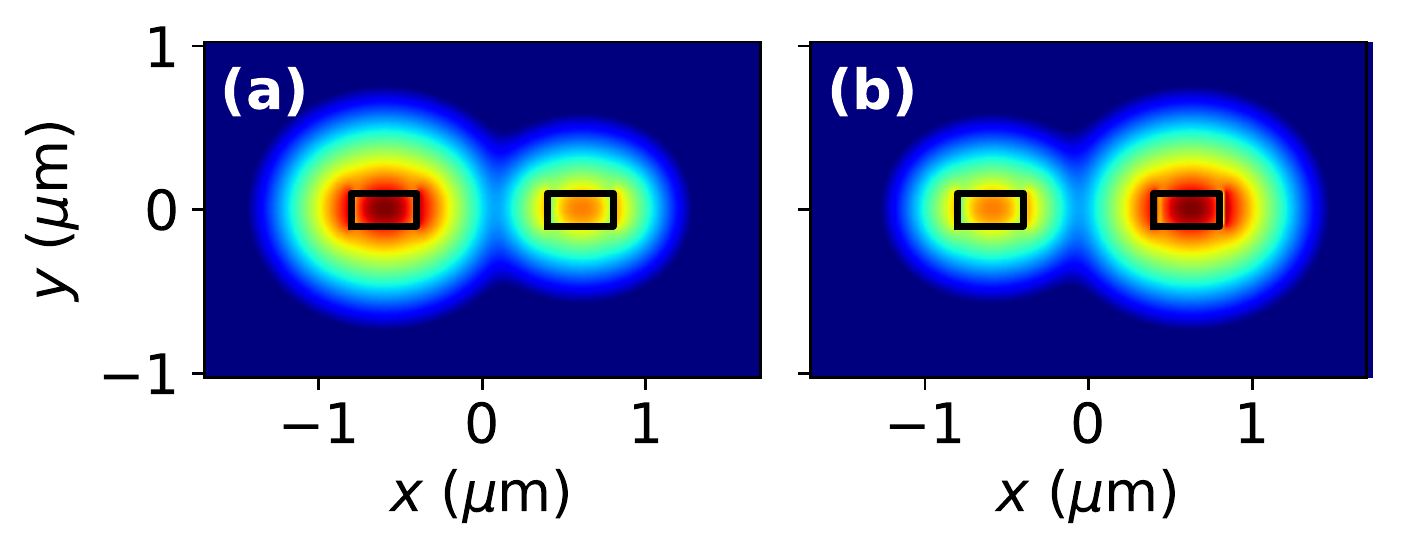}
  \caption{
    Distribution of the electric field component $E_x$
    in the broken-$\mathcal{PT}$-symmetric regime ($\gamma=8 \times 10^{-4}$).
    Distributions of $\abs{E_x}$ for (a) ``loss'' and (b) ``gain'' supermodes.
  }
  \label{fig:modes_above_ep}
\end{figure}

To characterize the system numerically we use VPIphotonics Mode Designer\texttrademark\ finite difference mode solver in the frequency domain \cite{ref:vpimd}.
We take parameters of the waveguide coupler as $g=0.8$ $\mu$m, $h=0.2$ $\mu$m, $n_\mathrm{cl}=1.444$, and $n_\mathrm{co}=3.478$ in order to limit the number of system's modes.
Refractive indices of the cladding and core correspond to those of $\text{SiO}_2$ and $\text{Si}$ at the wavelength $1.55~\mu\rm{m}$.
Then the coupler has only two quasi-TE supermodes at this wavelength.
The modes are visualized in Figs.~\ref{fig:modes_below_ep}~and~\ref{fig:modes_above_ep}.
In $\mathcal{PT}$-symmetric state, both the first and the second supermodes have symmetric distribution of the magnitude of the electric field $\abs{E_x}$ over the loss and gain waveguides ensuring a balance of the gain and loss [Figs.~\ref{fig:modes_below_ep}(a) and (b)].
The modes can be associated with the eigenvalues of the scattering matrix, which are known to be unimodular $\abs{s_{1,2}}=1$ and correspond to propagating waves of the form $s_{1,2}=\exp(-i\beta_{1,2}z)$.
Since in the Hermitian limit $\gamma=0$ the fields of the supermodes become real possessing even and odd symmetry, we call the supermodes ``even'' and ``odd'' in quotes for convenience.

In $\mathcal{PT}$-symmetry-broken regime, the fields of the supermodes have a completely different behavior.
According to Fig.~\ref{fig:modes_above_ep} the field is concentrated either in the loss or gain waveguide.
Hence, the supermodes can be named ``loss'' and ``gain'' modes.
In this case the supermodes are mirror reflections of each other with respect to the plane $x=0$.
The amplitude of the ``loss'' (``gain'') mode decreases (increases) during propagation in accordance with the known properties of the eigenvalues of the scattering matrix in the $\mathcal{PT}$-symmetry-broken state: $\abs{s_{1}}=1/\abs{s_{2}}$.

\begin{figure}[t!b!]
  \centering
  \includegraphics[width=\linewidth]{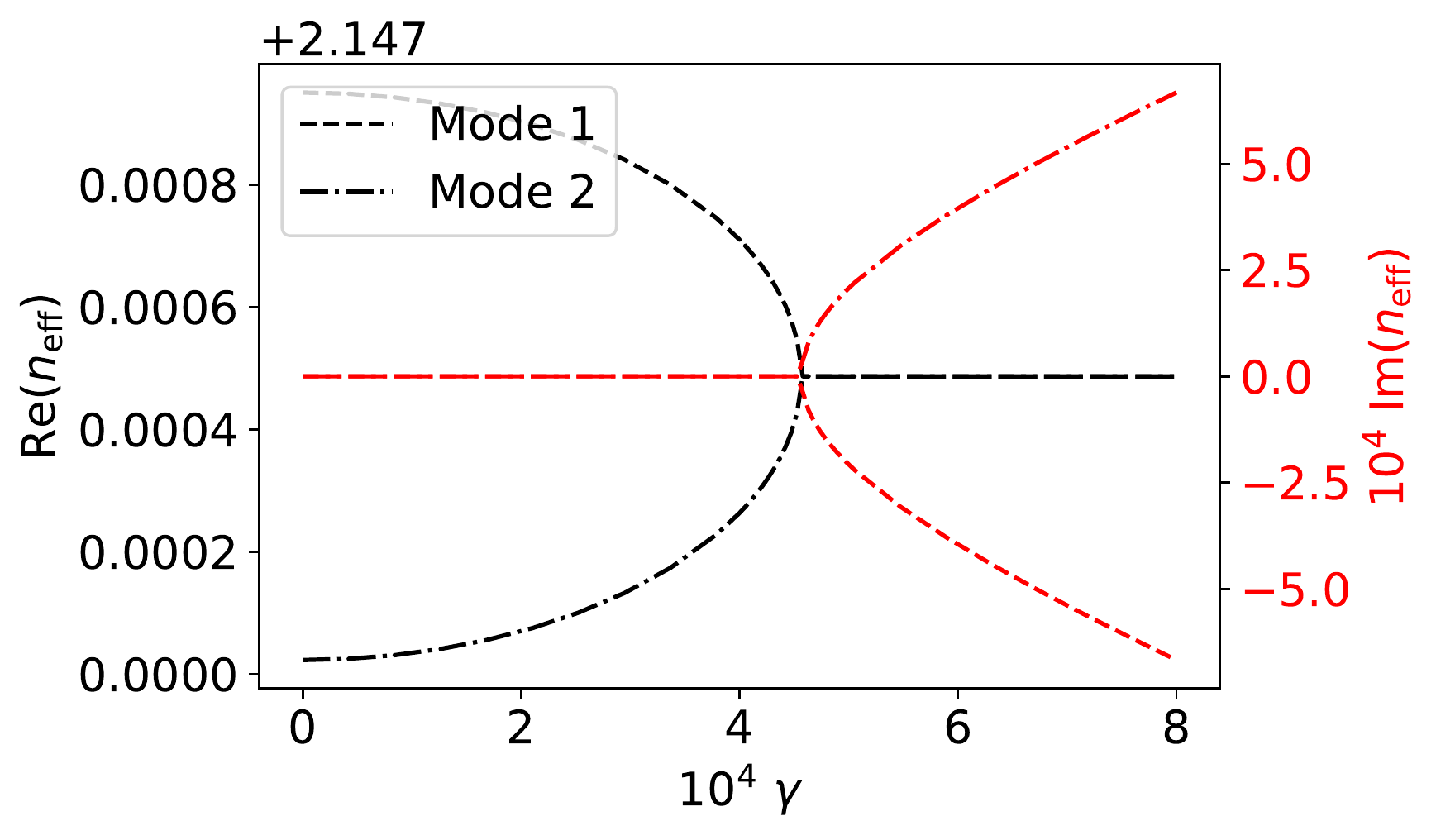}
  \caption{Effective mode indices versus the non-Hermiticity parameter $\gamma$.
  Black curves correspond to the real parts of the effective indices.
  Red curves correspond to the imaginary parts of the effective indices.
  Dashed (black and red) curves are related to the first supermode whereas dot-dashed ones related to the second supermode.}
  \label{fig:neff_3d}
\end{figure}

Transition from the $\mathcal{PT}$- to non-$\mathcal{PT}$-symmetric state occurs when varying some system's parameter. The transition is observed in the modal effective index of the coupled waveguides $n_{\rm eff} = {\rm Re}(n_{\rm eff}) + \ii~{\rm Im}(n_{\rm eff})$.
When increasing the gain/loss parameter $\gamma$ the system passes through the regime of propagation ($\mathcal{PT}$-symmetric state) for two non-decaying supermodes to the regime of decay/amplification ($\mathcal{PT}$-symmetry-broken state) for the modes with the refractive indices $n_{\rm eff} = {\rm Re}(n_{\rm eff}) \pm \ii~{\rm Im}(n_{\rm eff})$.
The curves in Fig.~\ref{fig:neff_3d} demonstrate this behavior. The non-$\mathcal{PT}$-symmetric phase emerges at the EP around $\gamma_\text{EP} = 4.21 \times 10^{-4}$.

\begin{figure}[t!b!]
  \centering
  \includegraphics[width=0.9\linewidth]{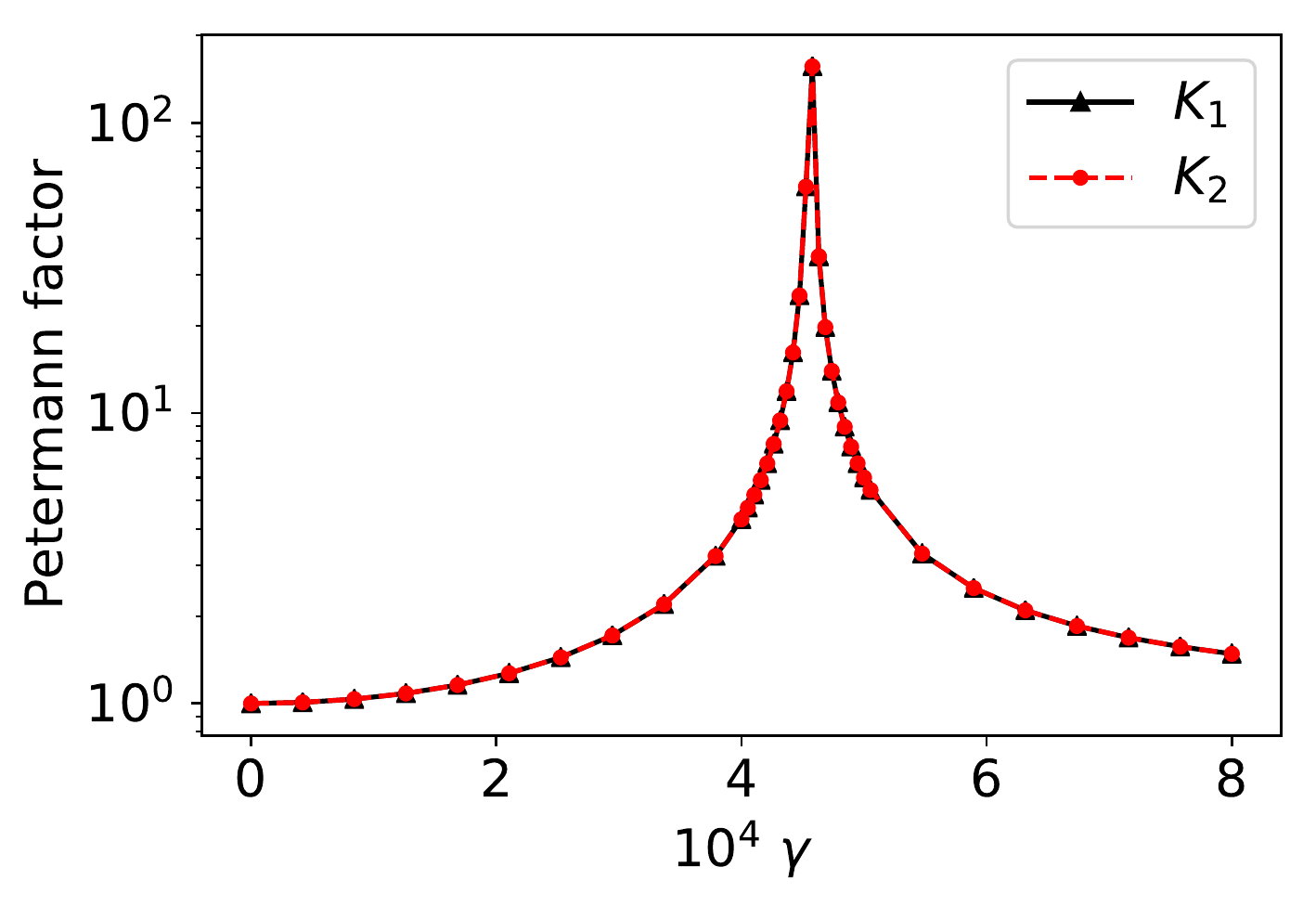}
  \caption{Petermann factors $K_1$ and $K_2$ respectively for the first and second supermodes of the $\mathcal{PT}$-symmetric coupled waveguides as functions of the non-Hermiticity parameter $\gamma$.
  Parameters of the waveguide coupler: $g=0.8$ $\mu$m, $h=0.2$ $\mu$m, $n_\mathrm{cl}=1.444$,
  and $n_\mathrm{co}=3.478$.
  }
  \label{fig:petermann}
\end{figure}

The Petermann factor for the supermodes in the coupled-$\mathcal{PT}$-symmetric waveguides depends on the non-Hermiticity parameter $\gamma$.
One can see in Fig.~\ref{fig:petermann} that the Petermann factors $K_{1,2}$ almost coincide for both supermodes.
When $\gamma$ approaches $\gamma_\text{EP}$, $K_{1,2}$ become singular.
This singularity might be considered as a consequence of the degeneracy of the modes of the $\mathcal{PT}$-symmetric system at the EP,
but a thorough analysis in Ref.~\cite{Pick2017} demonstrates that the peak value should be finite.
Similar result for the Petermann factor in $\mathcal{PT}$-symetric system was observed also in Ref.~\cite{ref:yoo2011}.

\begin{figure}[t!b!]
  \includegraphics[width=\linewidth]{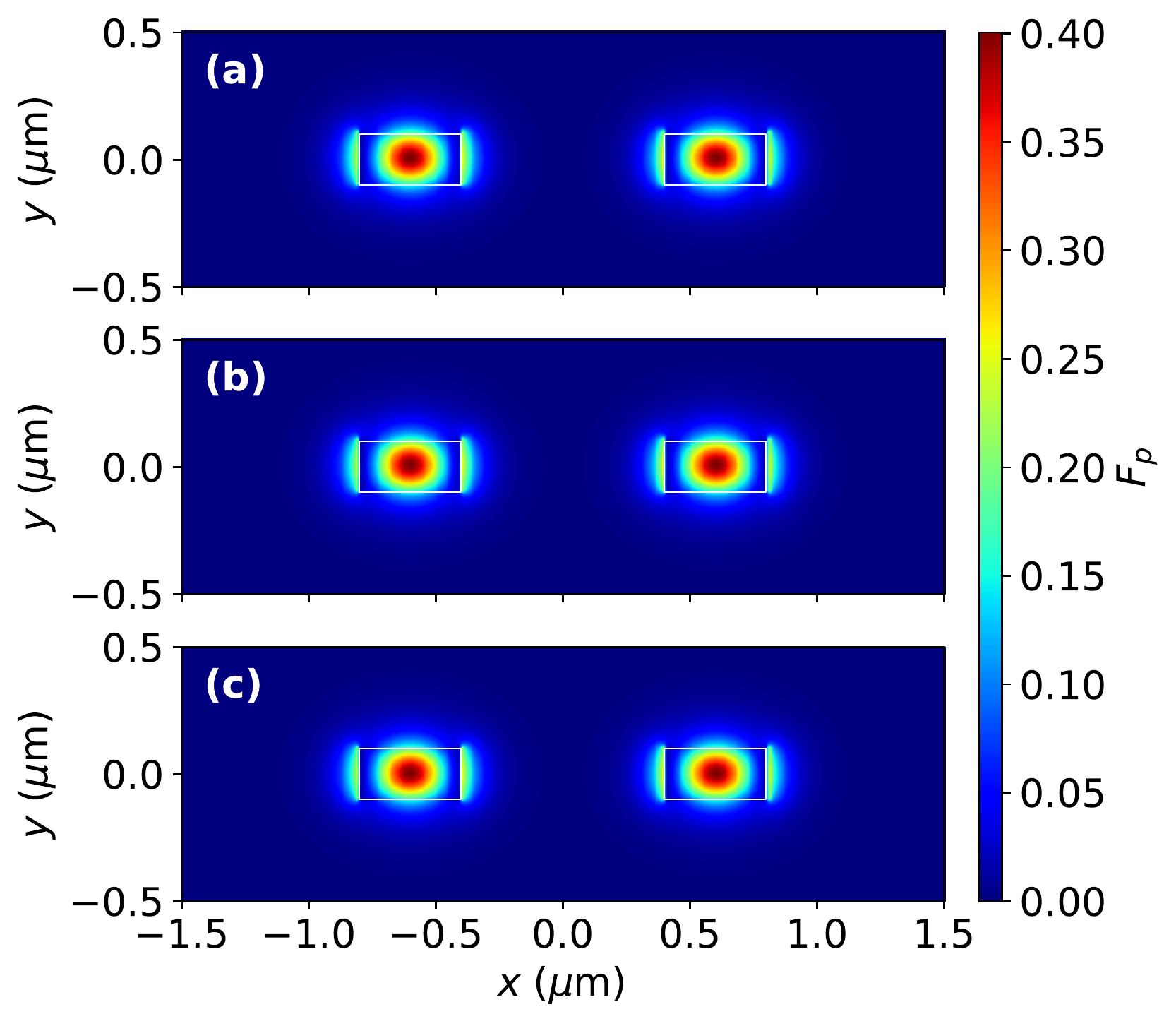}
  \caption{Purcell factor distribution in the plane ($x$, $y$) (a) for the Hermitian system characterized by $\gamma = 0$, (b) in the $\mathcal{PT}$-symmetric phase ($\gamma=3.5\times 10^{-4}$), (c) in the broken-$\mathcal{PT}$-symmetric state ($\gamma=8\times 10^{-4}$).
  Parameters of the waveguide coupler: $g=0.8$ $\mu$m, $h=0.2$ $\mu$m, $n_\mathrm{cl}=1.444$,
  and $n_\mathrm{co}=3.478$.
  }
  \label{fig:Fp_in_plane}
\end{figure}

While bearing in mind the theory developed in the previous section, we shall explore the Purcell factor $F_p$ as an enhancement factor of the spontaneous emission rate coupled to the pair of TE-like modes computed in Section II.
According to Eq.~\eqref{eq:Fpurcell}, the Purcell factor is defined by the fields of the reciprocal modes at the dipole position ($x_0$, $y_0$, $z_0\approx z_1 \approx z_n$).
In Fig. \ref{fig:Fp_in_plane}, we demonstrate the Purcell factor for an $x$-oriented dipoles as a function of
$x_0$~and~$y_0$ for different values of parameter $\gamma$ (imaginary part of the Gain waveguide refractive index $n_r$).

One can see in Fig.~\ref{fig:Fp_in_plane} that the modal Purcell factor is symmetric in
(a) Hermitian regime as well as in (b) $\mathcal{PT}$-symmetric and (c) $\mathcal{PT}$-symmetry broken regimes.
The Purcell factor $F_p$ is less than 1 taking a maximum value of approximately 0.4 in centers of the waveguides.

\begin{figure}[t!b!]
  \centering
  \includegraphics[width=0.9\linewidth]{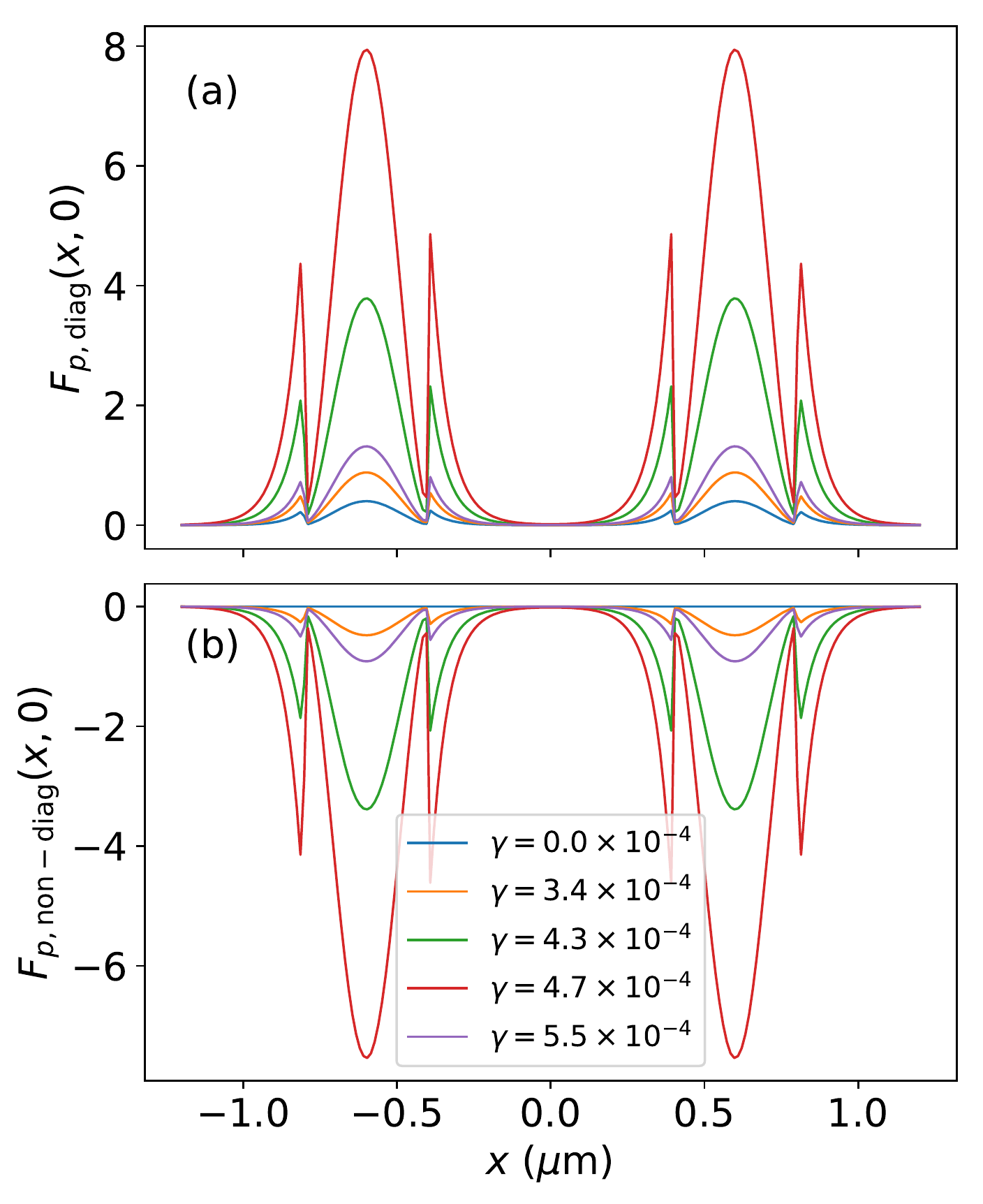}
  \caption{Distribution of the Purcell factor (a) diagonal and (b) non-diagonal terms depending
  on the emitter position $x_0$ at $y_0=0$ for different values of $\gamma$. Parameters of the coupled waveguide are given in the caption of Fig. \ref{fig:Fp_in_plane}.}
  \label{fig:Fp_gamma_y_is_0}
\end{figure}

According to Fig. \ref{fig:Fp_gamma_y_is_0} diagonal and non-diagonal terms have opposite signs and close absolute values. This explains small values of the modal Purcell factor in spite of the enhancement of $F_{diag}$ and $F_{non-diag}$ and their divergence at the EP.

\begin{figure}[t!b!]
  \centering
  \includegraphics[width=\linewidth]{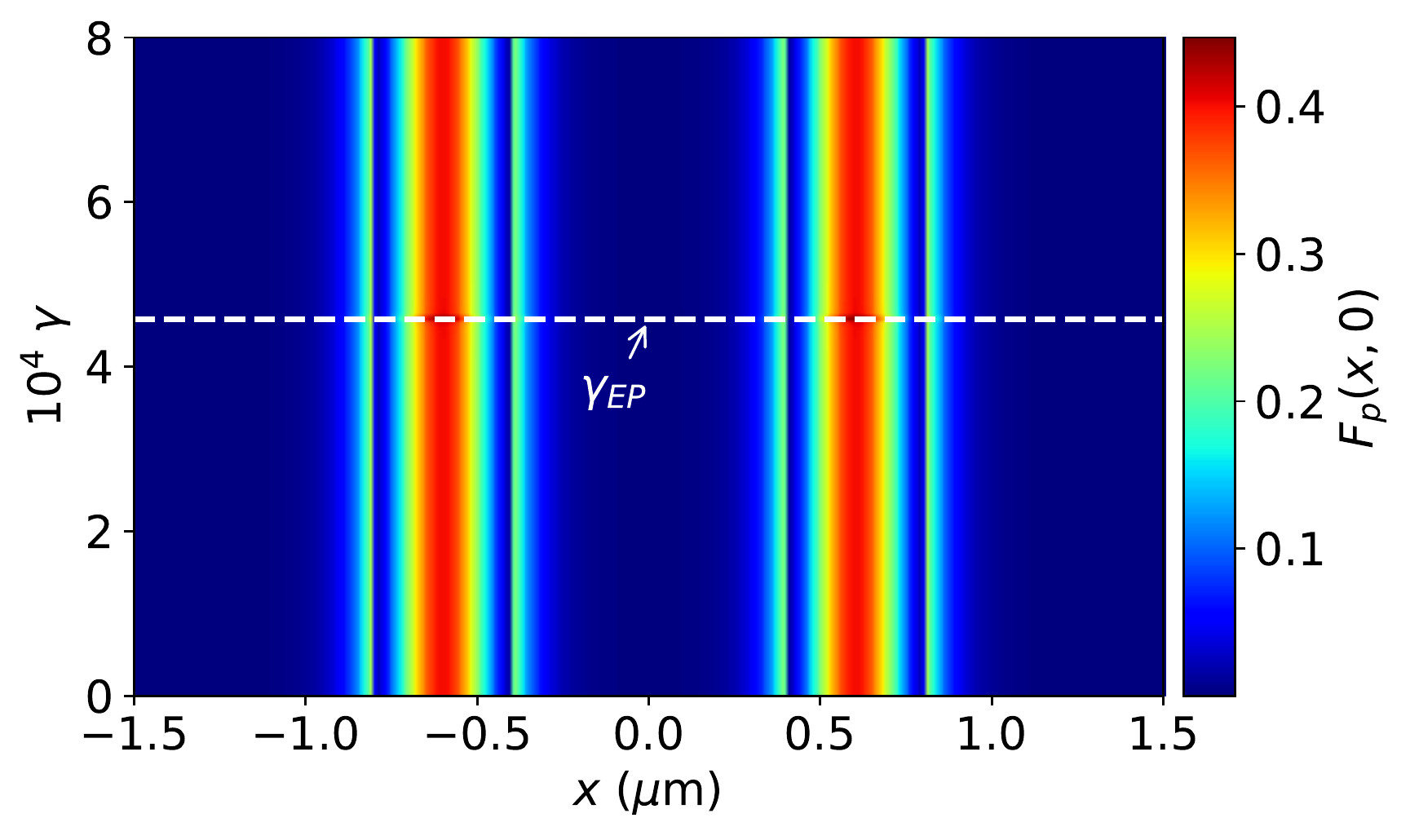}
  \caption{Distribution of the Purcell factor at the line
  $y=0$ as function of the emitter position $x$ and non-Hermiticity parameter~$\gamma$. Parameters of the coupled waveguide are given in the caption of Fig. \ref{fig:Fp_in_plane}. }
  \label{fig:Fp_x_gamma}
\end{figure}

Such a behavior well agrees with the result obtained in Section \ref{sec:method}
using the coupled-mode theory, namely, the numerically observed distribution of the modal Purcell factor is
similar in Hermitian,
$\mathcal{PT}$-symmetric, and $\mathcal{PT}$-symmetry broken regime. Independence of the non-Hermiticity parameter $\gamma$ including the exceptional point $\gamma_{EP}$ demonstrated in Fig. \ref{fig:Fp_x_gamma} also confirms the analytical predictions given by Eqns. \eqref{eq:Fp_sym} and \eqref{eq:Fp_broken}.

\begin{figure}[t!b!]
  \centering
  \includegraphics[width=\linewidth]{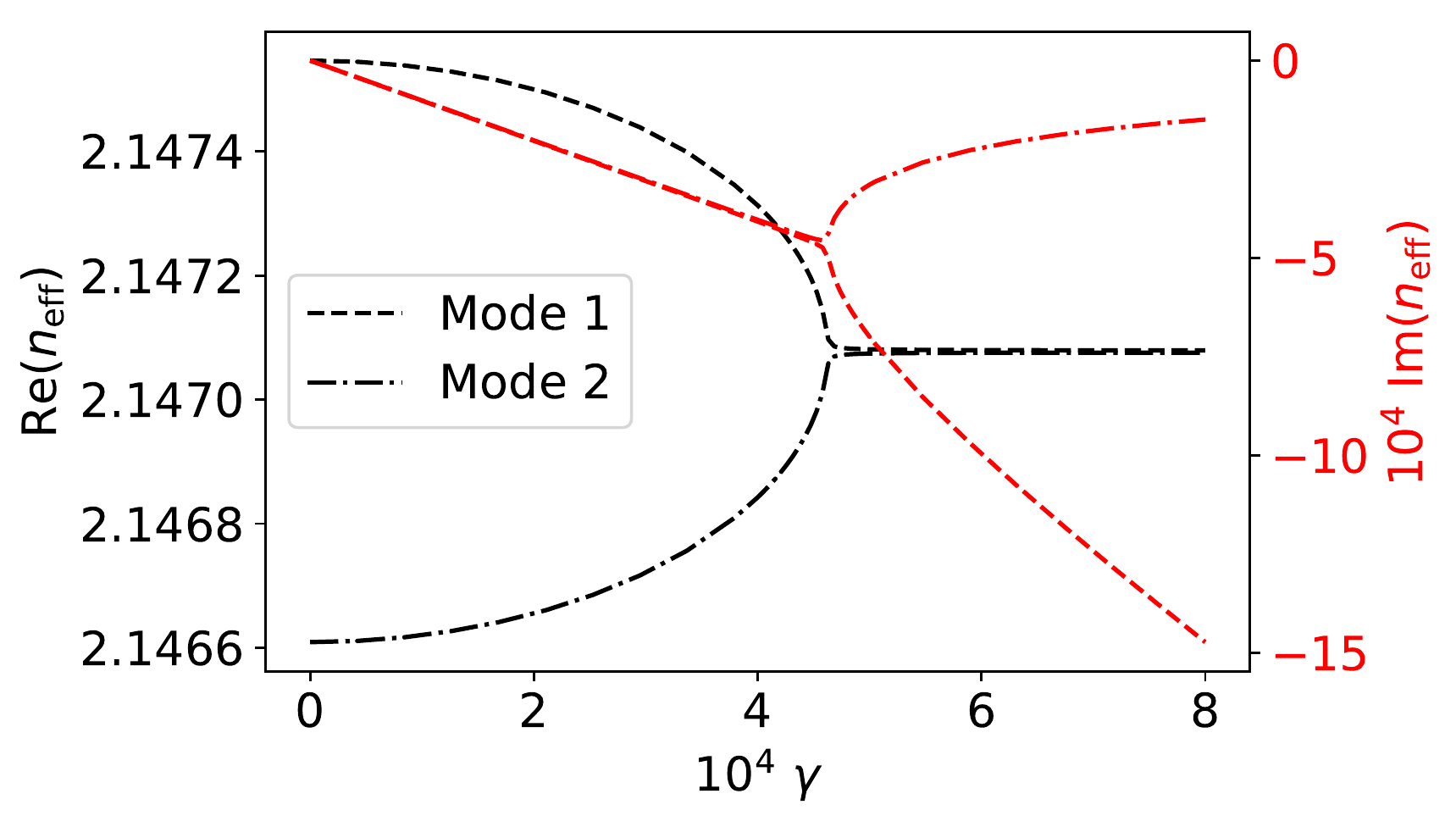}
  \caption{Effective mode indices for the passive coupler versus the non-Hermiticity parameter $\gamma$.
  Black curves correspond to real parts of the effective indices.
  Red curves correspond to imaginary parts.
  Dashed (black and red) curves are related to the first supermode whereas dot-dashed ones related to the second supermode. Parameters of the passive waveguide coupler: $g=0.8$ $\mu$m, $h=0.2$ $\mu$m, $n_\mathrm{cl}=1.444$,
  and $n_\mathrm{co}=3.478$.}
  \label{fig:neff_3d_passive}
\end{figure}

It is known that phase transition can occur also in entirely passive couplers, where the channels being either lossy or lossless.
The $\mathcal{PT}$ symmetry then is not exact \cite{ref:guo2009}.
We study a passive coupler with the same geometry as the coupler described previously in this paper.
In the passive coupler, the Gain waveguide is substituted with the lossless waveguide.
Imaginary part of the refractive index of the lossy waveguide is chosen to be $-2\gamma$.
For such a choice of parameters,
the phase transition in the passive coupler occurs at the same point as that in the original $\mathcal{PT}$-symmetric coupler.
This can be observed in Fig. \ref{fig:neff_3d_passive}.

\begin{figure}[t!b!]
  \centering
  \includegraphics[width=0.9\linewidth]{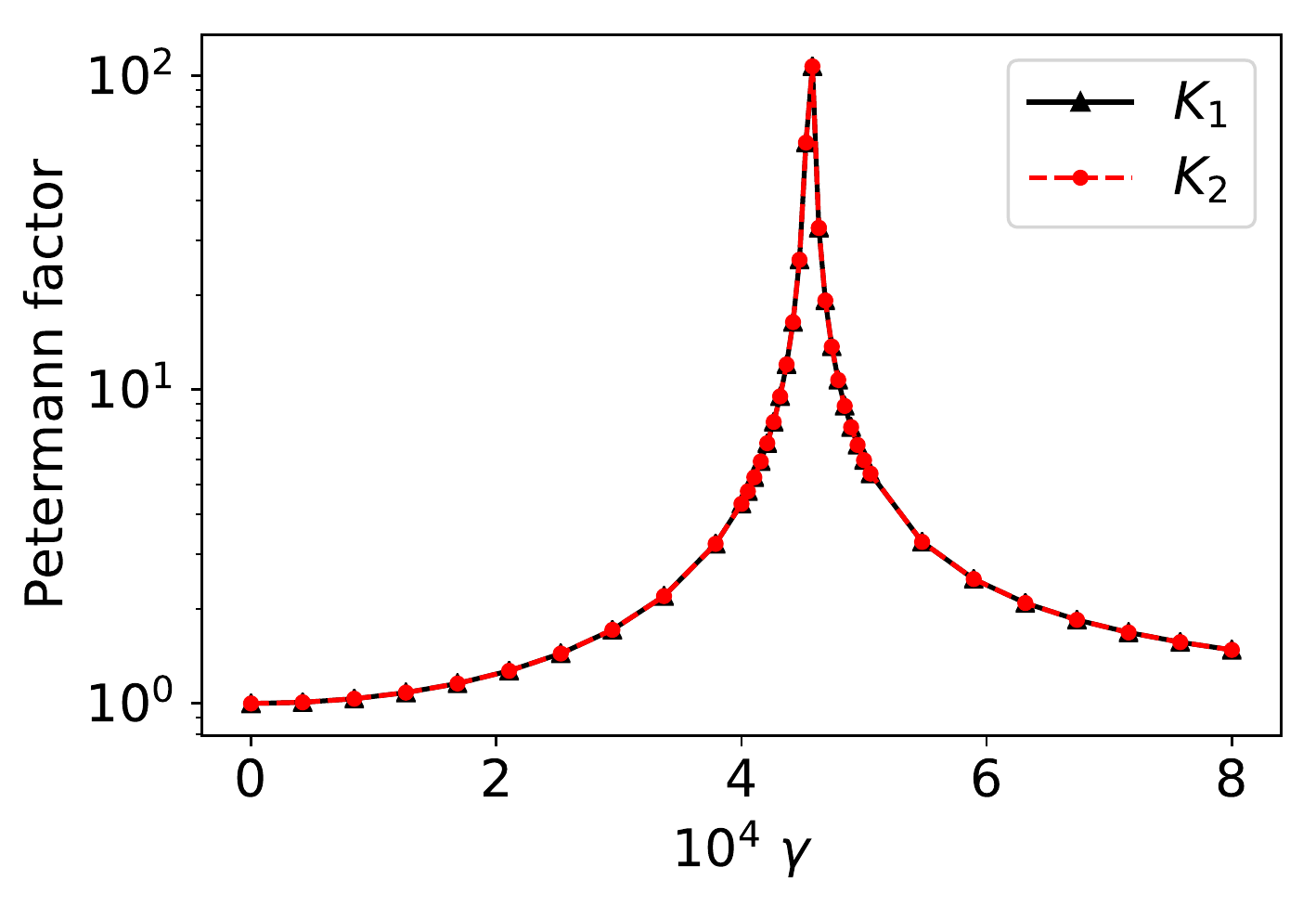}
  \caption{Petermann factors $K_1$ and $K_2$ respectively for the first and second supermodes of the passive coupler
   as functions of the non-Hermiticity parameter $\gamma$. Parameters of the guiding system are given in the caption of Fig. \ref{fig:neff_3d_passive}.}
  \label{fig:petermann_passive}
\end{figure}

\begin{figure}[t!b!]
  \includegraphics[width=\linewidth]{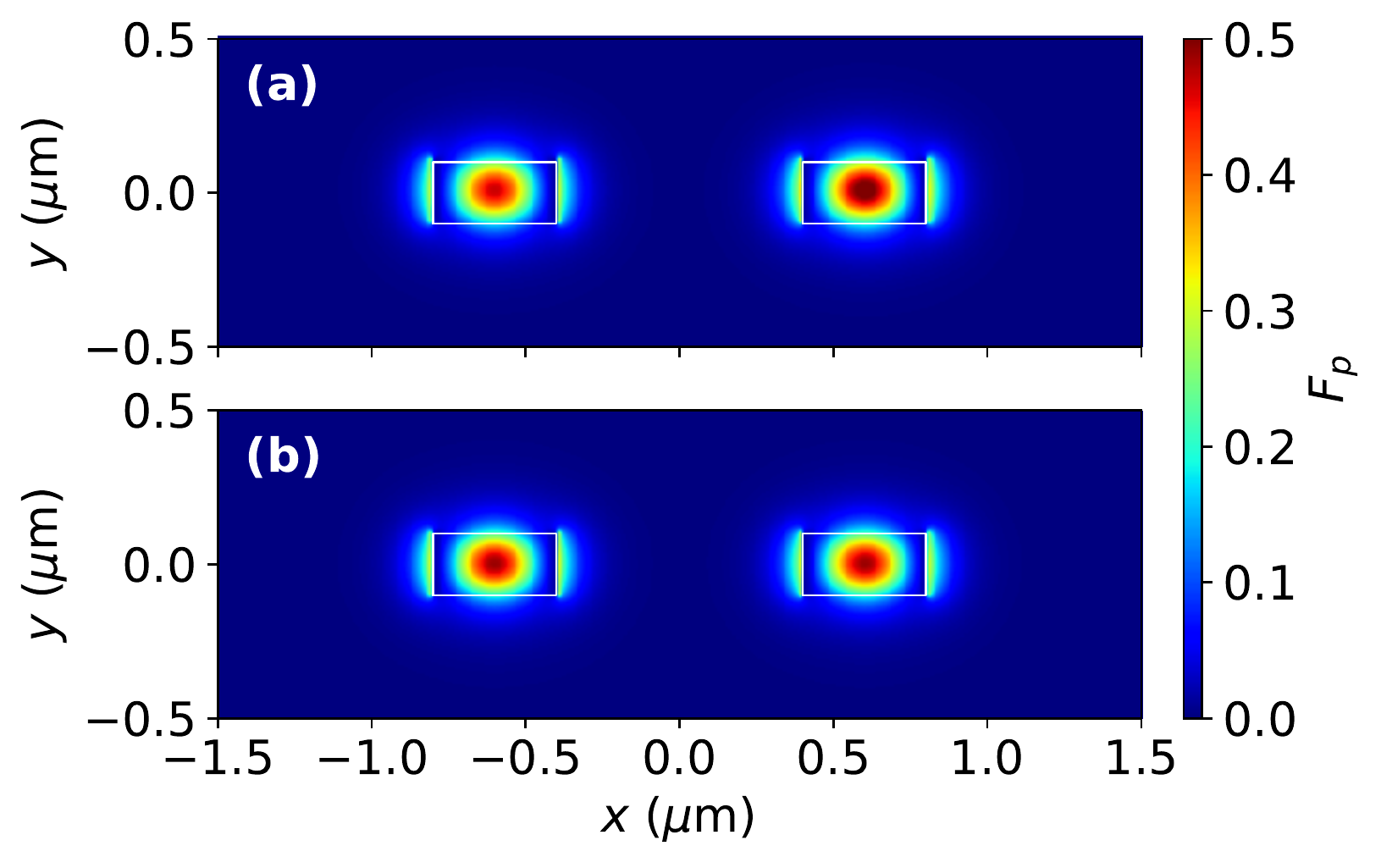}
  \caption{Purcell factor distribution in the plane ($x$, $y$)
    (a) below phase transition ($\gamma=3.5\times 10^{-4}$),
    (b) above the phase transition ($\gamma=8\times 10^{-4}$).
    Parameters of the guiding system are given in the caption of Fig. \ref{fig:neff_3d_passive}.
  }
  \label{fig:Fp_in_plane_passive}
\end{figure}

\begin{figure}[t!b!]
  \centering
  \includegraphics[width=\linewidth]{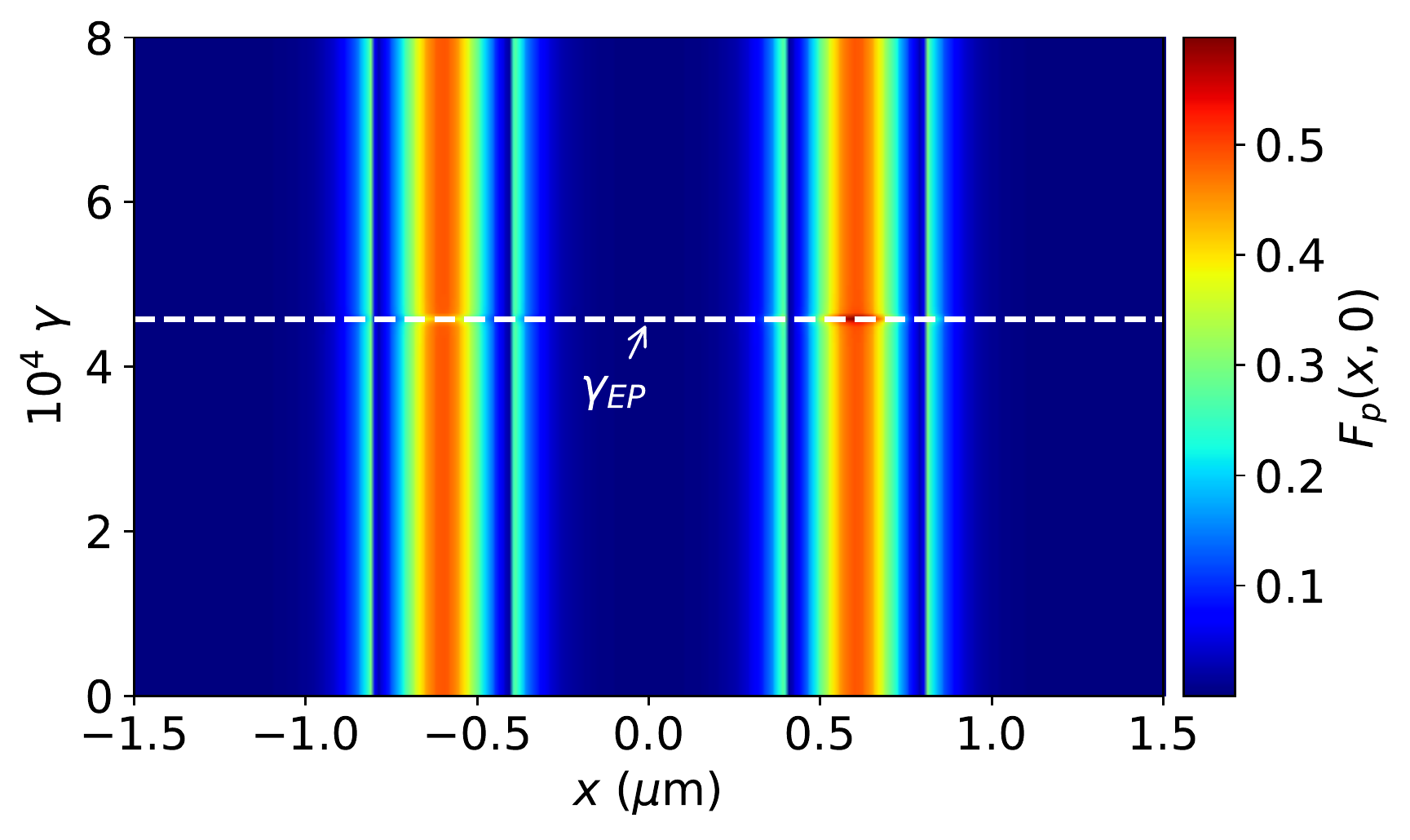}
  \caption{Distribution of the Purcell factor at the line
    $y=0$ as function of emitter position $x$ and non-Hermiticity parameter~$\gamma$.
    Parameters of the guiding system are given in the caption of Fig. \ref{fig:neff_3d_passive}.}
  \label{fig:Fp_x_gamma_passive}
\end{figure}

The Petermann factor is resonant at the exceptional point in the passive system as well (see Fig.~\ref{fig:petermann_passive}, and the modal Purcell factor in analogy with true $\mathcal{PT}$-symmetric system
shows no dependence on the non-Hermiticity parameter $\gamma$ as confirmed by
Figs.~\ref{fig:Fp_in_plane_passive}~and~\ref{fig:Fp_x_gamma_passive}.

We have verified results for the modal Purcell factor in the passive system by finite-difference time-domain (FDTD) simulations.
We have investigated Purcell enhancement for an $x$-polarized dipole source placed in the center of the lossless waveguide at different values of $\gamma$.
In full agreement with results obtained using reciprocity approach we have revealed almost no change in the Purcell factor in comparison to that in Hermitian system.
FDTD simulations were performed using an open-source software package~\cite{ref:oskooi2010}.

\section{Summary}
In this paper, we have reported on the investigation of the spontaneous emission rate enhancement for a point-source emitter in $\mathcal{PT}$-symmetric system of coupled waveguides.
We have generalized the reciprocity technique proposed in Ref.~\cite{ref:schulz2018} taking into account the non-orthogonality of modes of the $\mathcal{PT}$-symmetric system.
We have revealed analytically using the coupled-mode approach that the Purcell factor for $\mathcal{PT}$-symmetric system of coupled waveguides does not depend on the non-Hermiticity taking close values for Hermitian and $\mathcal{PT}$-symmetric systems.
Even at the exceptional point, where the Petermann factor diverges due to the modes
self-orthogonality, the modal Purcell factor remains finite and almost coincides with
that for the Hermitian system.
Such a behavior of the Purcell factor is motivated by interplay of in-mode and cross-mode
terms, that diverge themselves at the EP, resulting in compensation of each other.
This result is supported with the general theory of spontaneous emission near exceptional points developed in Ref.~\cite{Pick2017}, where rigorous treatment of degeneracies shows that the Purcell factor even in gain-assisted systems remains finite.


\section{Acknowledgements}
We acknowledge Sergei Mingaleev for valuable comments and VPIphotonics company for providing Mode Designer\texttrademark\ as mode solving software. A.K. acknowledge Israel Innovation Authority KAMIN program Grant no. 69073. F.M. and A.N. thank the Belarusian Republican
Foundation for Fundamental Research (Project No. F18R-021)

\appendix

\section{\label{sec:Appendix} Non-Hermitian reciprocity approach for a cavity}

Generally, a cavity causes reflection and transmission of the reciprocal mode $\ket{m_{-k, z_1}}$:
\begin{align}
  \ket{\psi_{2}(z_1)} =& B_{-k, z_1} \ket{-k,z_1} \ee^{\ii \beta_{k} z_1} \nonumber \\
  &+\sum_i B_{i, z_1} \ket{i,z_1} \ee^{-\ii \beta_{i} z_1},
  \nonumber \\
  \ket{\psi_{2}(z_n)} =& \sum_i B_{-i, z_n} \ket{-i,z_n} \ee^{\ii \beta_{i} z_n}.
  \label{eq:app_psi2zn}
\end{align}

Using the orthogonality condition
\eqref{eq:orth} and symmetry relations \eqref{eq:forward_backward} we obtain the inner products of the fields
\begin{subequations}
\begin{multline}
  \braket{\psi_1(z_1)}{\psi_2(z_1)} = \\
  \sum_{i}A_{i,z_1}B_{-k,z_1}\ee^{-\ii\beta_iz_1+\ii\beta_kz_1}
  \bra{i,z_1}\ket{-k,z_1}\\
  + \sum_{i, j}A_{i,z_1}B_{j,z_1}\ee^{-\ii\beta_iz_1-\ii\beta_jz_1}\bra{i,z_1}\ket{j,z_1}\\
  =-2A_{k,z_1}B_{-k,z_1}N_k +
  2\sum_{i}A_{i,z_1}B_{i,z_1}N_{i,z_1}\ee^{-2\ii\beta_iz_1}.
\end{multline}
\begin{multline}
\braket{\psi_2(z_1)}{\psi_1(z_1)} =\\
  \sum_{i}A_{i,z_1}B_{-k,z_1}\ee^{-\ii\beta_iz_1+\ii\beta_kz_1}
  \bra{-k,z_1}\ket{i,z_1}\\
  + \sum_{i, j}A_{i,z_1}B_{j,z_1}\ee^{-\ii\beta_iz_1-\ii\beta_jz_1}
  \bra{i,z_1}\ket{j,z_1}\\
  =2A_{k,z_1}B_{-k,z_1}N_{k,z_1} +
  2\sum_{i}A_{i,z_1}B_{i,z_1}N_{i,z_1}\ee^{-2\ii\beta_iz_1}.
\end{multline}
\begin{multline}
  \braket{\psi_1(z_n)}{\psi_2(z_n)}=\braket{\psi_2(z_n)}{\psi_1(z_n)}=\\
  \sum_{i, j}A_{-i,z_n}B_{-j,z_n}\ee^{\ii\beta_iz_n+\ii\beta_jz_n}
  \bra{-i,z_n}\ket{-j,z_n}\\
  =2\sum_{i}A_{-i,z_n}B_{-i,z_n}N_{i,z_n}\ee^{2\ii\beta_iz_n}.
\end{multline}
\label{eq:rhs_recipr}
\end{subequations}

By substituting these equations into the reciprocity theorem (\ref{eq:reciprocity1}),
we again derive Eq. (\ref{eq:minus4ab}).

\bibliography{references}

\end{document}